# Sol-gel synthesis of $Li_3VO_4$/C composites as anode materials for lithium-ion batteries

E. Thauer[1], G.S. Zakharova[2], S.A. Wegener[1], Q. Zhu[3], R. Klingeler[1,4]

[1] Kirchhoff Institute of Physics, Heidelberg University, Heidelberg, Germany

[2] Institute of Solid State Chemistry, Ural Division, Russian Academy of Sciences, Yekaterinburg, Russia

[3] Institute of Material Science and Engineering, Wuhan University of Technology, Wuhan, PR China

[4] Centre for Advanced Materials, Heidelberg University, Heidelberg, Germany

**Abstract**

$Li_3VO_4$/C composites have been synthesized by a sol-gel method and post-annealing at 650 ºC for 1 h in $N_2$ flow using either tartaric acid, malic acid, or glucose as both chelating agents and carbon source. The presence of these organic additives crucially affects morphology and crystallite size of the final product. It is found that the electrochemical properties of $Li_3VO_4$/C as anode material for Li-ion batteries (LIBs) are influenced by the morphology, texture and carbon content of the material. When using carboxylic acids as carbon source composites with mesoporous structure and a high surface area are obtained that display an enhanced electrochemical activity. Initially, reversible capacity of about 400 mAh $g^{-1}$ is obtained. In contrast, $Li_3VO_4$/C synthesized with glucose outperforms in terms of cycling stability. It exhibits a discharge capacity of 299 mAh $g^{-1}$ after 100 cycles corresponding to an excellent capacity retention of 96 %. The favorable effect of carbon composites on the electrochemical performance of $Li_3VO_4$ is shown.

1. Introduction

An ideal Li-ion battery for commercial application has to meet a variety of requirements such as high energy density, high safety performance, a long lifetime and a good rate capability. One approach to achieve these goals is the development of advanced anode materials. The common commercial anode materials for Li-ion batteries, i.e., graphite, suffers from a low insertion potential coming along with security risks due to the formation of lithium dendrites and its moderate rate capability caused by its low Li$^+$ diffusion coefficient [1]. In general, three types of Li$^+$ storage mechanism may be distinguished: intercalation, conversion and alloying [2]. The key advantages of intercalation materials for commercial use are superior reversibility and good stability due to small volume changes during the insertion and extraction of Li-ions into a host lattice. Despite the development of successful strategies for improving the cycle stability of conversion and alloy-based anode materials made in recent years, they cannot compete with intercalation anodes [3]. Among these, especially the zero-strain material Li$_4$Ti$_5$O$_{12}$ has attracted much attention due to its excellent cycle stability and its high Li$^+$ insertion potential of 1.55 V [4]. Although beneficial in terms of safety the high operating voltage along with a low specific capacity of only 175 mAh g$^{-1}$ result in a low energy density. Thus, Li$_4$Ti$_5$O$_{12}$ finds proper use as anode material only in niche applications [5]. Li$_3$VO$_4$ is another promising intercalation material which was first studied by Li et al. as anode material for Li-ion batteries [6]. In contrast to Li$_4$Ti$_5$O$_{12}$ it exhibits a higher theoretical capacity of 397 mAh g$^{-1}$ corresponding to x = 2 in Li$_{3+x}$VO$_4$, which, that is comparable to that of graphite (372 mAh g$^{-1}$), and the insertion of Li-ions occurs at a safe but still low-potential window between 0.2 - 1 V [7]. Another advantage of Li$_3$VO$_4$ is its high mobility of Li-ions [8]. However, to become a high-performance anode material one has to overcome its low electronic conductivity resulting in large resistance polarization and poor rate capability. Composites of $Li_3VO_4$ with carbon-based materials has been demonstrated as an effective strategy to solve this problem [7,9–16]. A number of different carbon materials, such as carbon nanotubes [17,18], graphite [19,20], graphene [11,12,21,22] and graphene oxide [23,24], as well as various synthesis methods have been utilized to prepare $Li_3VO_4$/C composites. Besides promoting fast electron transfer, the carbon network can act as buffer of the volume changes and provides the structural stability of the electrode during cycling. Moreover, attributed to its stable chemical characteristics [25], carbon as surface layer can protect the material from side reactions at the electrode/electrolyte interface [26–28]. Surface coating is a promising approach to boost the structural stability of electrode materials and a hotspot research field for cathodes [29–32]. The electrochemical properties of $Li_3VO_4$/C are significantly influenced by the carbon component and the preparation method. For example, $Li_3VO_4$/C composite with a capacity of 245 mAh $g^{-1}$ after 50 cycles at 20 mA $g^{-1}$ were prepared via solid-state synthesis combined with chemical vapor deposition (CVD) coating by Shao et al. [28]. The *in-situ* carbon-encapsulated $Li_3VO_4$ nanoparticles synthesised via solid-state method by Zhang et al. [15] show a reversible capacity of 340 mAh $g^{-1}$ at a current density of 4 A $g^{-1}$ and a capacity retention of 80 % after 2000 cycles. Ni et al. report the hydrothermal synthesis of $Li_3VO_4$/C composite using citric acid as carbon source which delivers after 1000 cycles at 2 A $g^{-1}$ a specific discharge capacity of 422 mAh $g^{-1}$ [33]. Qin et al. [26] used electrospinning to prepare $Li_3VO_4$/C nanofibers which exhibits a capacity of 405 mAh $g^{-1}$ at 40 mA $g^{-1}$. Among different fabrication techniques, the sol–gel method provides a molecular level mixing of the starting materials which enhances chemical homogeneity of the final products. Additionally, low temperature of the preparation facilitates the formation of mesopores on the material surface, thereby improving the electrochemical properties. For example, carbon-coated $Li_3VO_4$ nanoparticles fabricated by citric acid-assisted sol-gel route achieved a high reversible capacity of 480 mAh g$^{-1}$ at 0.1 A g$^{-1}$ [34]. In the sol-gel process, the chelating agent takes part in hydrolysis and condensation reactions of metal-oxide precursor, forming M-O-M bridging bonds by the chelation between metal ions and polar functional groups of the chelating agent.

In the present study, three different organic additives, i.e., glucose, malic acid and tartaric acid, are used to synthesize $Li_3VO_4$/C composites through a facile sol-gel thermolysis method. Malic acid and tartaric acid, having carboxyl COOH-functional groups, belong to the acid chelating agents and tend to diminish the pH of the metal-oxide precursor solution to lower values. The use of tartaric and malic acid as a chelating agent for the synthesis of $Li_3VO_4$/C composites has not been reported so far. On the contrary, glucose has a ring structure with hydroxyl OH-groups and can be a chelating agent only in acidic ambient, turning into gluconic acid. In this work, in base medium, glucose serves as a capping agent to prevent the cations from agglomerating generating a highly viscous and stable sol [35]. Additionally, all these additives are carbon-rich materials which can be used as carbon sources to produce composite materials and enhance the crystallization process during the calcination as well. Here, we report the influence of different type of organic additives on the phase composition, morphology and electrochemical properties of the $Li_3VO_4$/C composites.

## 2. Experimental

### 2.1. Materials preparation

Ammonium metavanadate $NH_4VO_3$, lithium hydroxide monohydrate $LiOH \cdot H_2O$, glucose $C_6H_{12}O_6$, malic acid $C_4H_6O_5$, tartaric acid $C_4H_6O_6$ purchased from Sigma–Aldrich were used for the synthesis. First, 0.585 g $NH_4VO_3$ and 0.630 g $LiOH \cdot H_2O$ (with the molar ratio of $NH_4VO_3$ to $LiOH \cdot H_2O$ being 1:3) were dissolved under stirring in 50 ml distilled water to form a clear solution. Afterwards, glucose (with the molar ratio of glucose to vanadium metal of 1:1) was employed as carbon source and was added to the solution. To get a gel, the solution was heated at 60 °C under continuous stirring. The resulting gel was dried at 60 °C in air to get precursor, and later, calcinated at 650 °C for 1 h in $N_2$ flow to yield the $Li_3VO_4$/C composite. The as-prepared product is labeled as $Li_3VO_4$/C-G. For comparison, $Li_3VO_4$/C was produced with malic acid and tartaric acid at the same molar ratio of V : carboxylic acid = 1 : 1. These samples are denoted as $Li_3VO_4$/C-M and $Li_3VO_4$/C-T, respectively. The bare $Li_3VO_4$ sample was prepared by annealing of the $Li_3VO_4$/C-M composite at 650 °C for 1 h in air flow.

2.2. Materials characterization

X-ray powder diffraction (XRD) patterns were taken with a Bruker AXS D8 Advance Eco using Cu Kα radiation (λ = 1.540 Å). The morphology of the samples was investigated by a ZEISS Leo 1530 scanning electron microscope (SEM) and a JEOL JEM 2100 transmission electron microscope (TEM). Thermogravimetric analysis (TG-DSC-MS) with a heating rate of 10 K·min$^{-1}$ under flowing air was carried out using STA 449 $F_3$ Jupiter thermoanalyzer (Netzsch) coupled with a QMS 403 mass spectrometer. The examination of the carbon content was carried out by chemical analysis (CA) using Vario MICRO Cubes (Elementar). Raman spectra were measured with a Renishaw in Via Reflex spectrometer at a laser wavelength of 532 nm. Nitrogen sorption isotherms were determined on a Micromeritics Gemini VII 2390 Surface Area Analyzer. The specific surface area, pore size distribution and pore volumes were obtained by means of the Brunauer-Emmett-Teller (BET) method and the Barrett-Joyner-Halenda model from the desorption branches of the isotherms.

*2*.3. Electrochemical Measurements

The electrochemical measurements were carried out with a VMP3 potentiostat (Bio-Logic SAS) at 25 °C using Swagelok-type cells (see [36]). For the preparation of the working electrodes 80 wt% Li$_3$VO$_4$/C, 15 wt% carbon black (Super C65, Timcal) and 5 wt% polyvinylidene fluoride binder (PVDF, Solvay Plastics) dissolved in N-methyl-2-pyrrolidone (NMP, Sigma-Aldrich) were mixed and stirred for at least 12 h. The spreadable slurry obtained by evaporating most of the NMP was applied on circular Cu meshes with a diameter of 10 mm. Afterwards, the electrodes were dried at 80 °C under vacuum, mechanically pressed with 8 MPa and dried again. The loading density was about 4 mg cm$^{-2}$. A lithium metal foil (Alfa Aesar) pressed on a nickel disk served as counter electrode. Two layers of glass fiber (Whatman GF/D) that was soaked with 200 μl electrolyte (Merck Electrolyte LP30), a 1 M LiPF$_6$ salt solution in 1:1 ethylene carbonate and dimethyl carbonate were used as separator. The cells were assembled in a glove box under argon atmosphere (O$_2$/H$_2$O < 5 ppm). For *ex-situ* XRD measurements the cells were disassembled in a glovebox and the electrodes were washed in DMC. The XRD measurements were performed using an airtight sample carrier. The calculation of the specific capacity is based on the total mass weight of the composites Li$_3$VO$_4$/C or of bare Li$_3$VO$_4$, respectively.

## 3. Results and Discussion

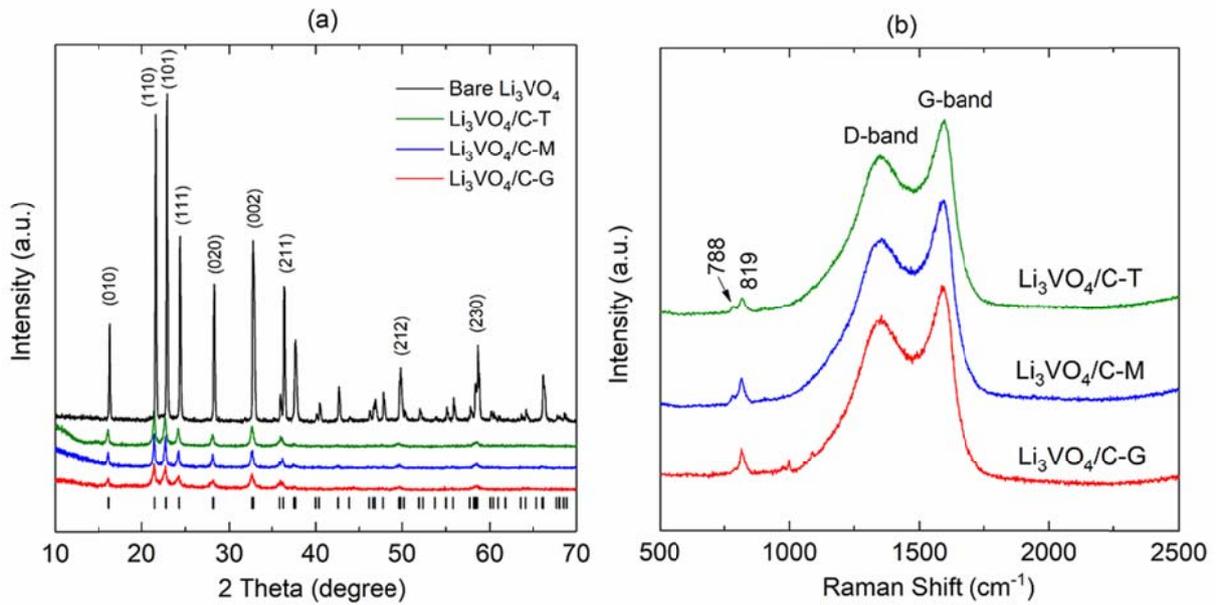

Fig.1. (a) XRD patterns of the $Li_3VO_4$/C composites synthesized with glucose (G), malic acid (M), and tartaric acid (T) as carbon source as well as of bare $Li_3VO_4$. Black bars mark the Bragg peak position of $Li_3VO_4$ according to COD #1528868 [37]. (b) Raman spectra of the $Li_3VO_4$/C composites.

The XRD patterns of the $Li_3VO_4$/C composites shown in Fig. 1a confirm that by heating the precursors up to 650 °C composites based on $Li_3VO_4$ are formed. All diffraction peaks can be assigned to $Li_3VO_4$ with an orthorhombic structure according to the reference COD-1528868 [37]. As shown in Table 1, the lattice parameters of $Li_3VO_4$ in the composites are in a good agreement with reference data. The estimated average crystallite size of $Li_3VO_4$ calculated by Scherrer's formula based on the (110), (101) and (111) crystal planes indicates diameters of 30 – 50 nm.

Table 1. Lattice parameters from Rietveld refinement, average crystallite size from analyzing the XRD peak width of the (110), (101) and (111) reflections, carbon content, and BET surface area for $Li_3VO_4$/C composites. The last two lines show parameters of $Li_3VO_4$ studied here and from the literature [37].

| Sample | Lattice parameters | $D_{XRD}$ | Carbon content (wt%) | $S_{BET}$ |
| --- | --- | --- | --- | --- |

|  | a (Å) | b (Å) | c (Å) | (nm) | TGA | CA | (m$^2$ g$^{-1}$) |
|---|---|---|---|---|---|---|---|
| Li$_3$VO$_4$/C-G | 6.350(1) | 5.460(2) | 4.959(4) | 25(9) | 25.6 | 18.0(5) | 12 |
| Li$_3$VO$_4$/C-M | 6.346(3) | 5.462(4) | 4.964(2) | 41(7) | 13.7 | 10.7(5) | 42 |
| Li$_3$VO$_4$/C-T | 6.341(8) | 5.461(9) | 4.973(8) | 34(8) | 14.1 | 9.6(5) | 32 |
| Bare Li$_3$VO$_4$ | 6.329(1) | 5.448(1) | 4.949(1) | 61(5) |  | 0 |  |
| Li$_3$VO$_4$ [COD-1528868] | 6.3259 | 5.446 | 4.9469 |  |  |  |  |

To confirm and to quantify the presence of carbon in the Li$_3$VO$_4$/C composites, Raman spectra and DSC-TG-MS curves were measured. The Raman spectra of all samples show two broad peaks at about 1340 and 1593 cm$^{-1}$ (Fig. 1b). The Raman-active E$_{2g}$ mode at 1593 cm$^{-1}$ (G-band) is characteristic of sp$^2$-hybridized carbon atoms resulting from in-plane vibrations, whereas the D-band at around 1340 cm$^{-1}$ can be attributed to the presence of structural defects or disorder [38]. The ratio of the maximal intensities of the D- and G-band (I$_D$/I$_G$) was calculated as 0.87, 0.82, 0.82 for Li$_3$VO$_4$/C-G, Li$_3$VO$_4$/C-T, and Li$_3$VO$_4$/C-M composites, respectively, indicating a small fraction of sp$^3$-hybridized carbon atoms [39]. A high graphitization degree is supposed to enhance the electron transport and thus the electrochemical cycling performance [22]. The peaks located around 788 and 819 cm$^{-1}$ are attributed to the asymmetric and symmetric stretching vibrations of VO$_4$-tetrahedra in Li$_3$VO$_4$, respectively [17,40]. These results confirm the successful preparation of Li$_3$VO$_4$/C composites.

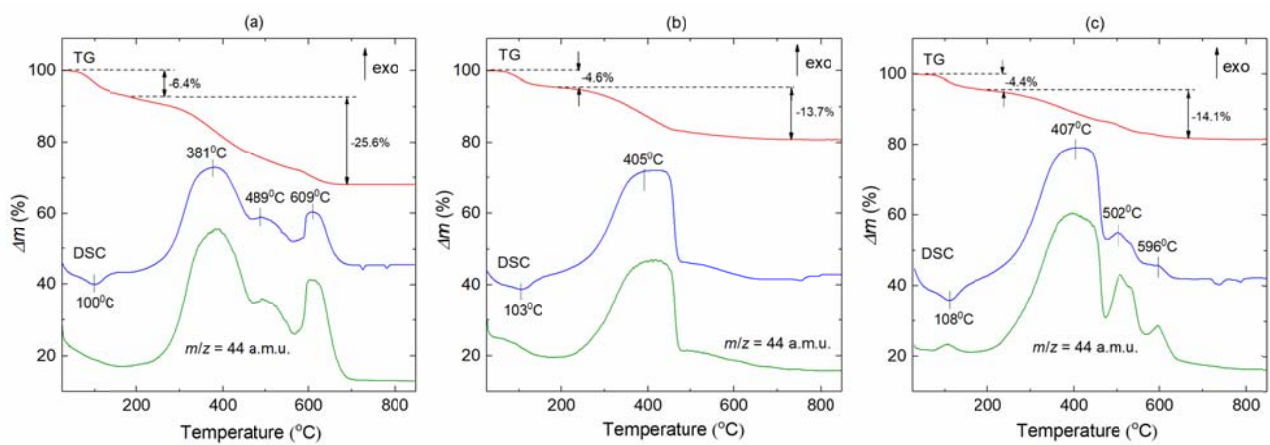

Fig. 2. Thermogravimetric (red), DSC (blue), and mass spectroscopy (green) curves of (a) $Li_3VO_4$/C-G, (b) $Li_3VO_4$/C-M, and (c) $Li_3VO_4$/C-T composites.

In order to determine the carbon content of $Li_3VO_4$/C composites, TG-DSC-MS measurements of as-prepared materials were carried out under air flow. It can be observed that several steps emerge from the TG-curves of the $Li_3VO_4$/C composites (Fig. 2). In all samples, the first step starts from about 70 ºC and ends at 200 ºC. It can be mainly attributed to the loss of water absorbed on the material's surface. The process is accompanied by a weak and broad endothermic peak centered at about 104 ºC. Additionally, the MS-curve of the $Li_3VO_4$/C-T composite (ion current versus temperature) also implies a weight loss in the range of 70 – 170 ºC ascribed to evaporation of surface-adsorbed gaseous carbon dioxide (m/z = 44 a.m.u.). The next weight loss from 250 ºC to 700 ºC is caused by oxidation of the carbon component of $Li_3VO_4$/C composites and the emission of $CO_2$ gases as also corroborated by analysis of the MS curves. This process is accompanied by a broad and complex exothermal peak. The MS curves exhibit maxima at the same temperatures where exothermic peaks appear. For $Li_3VO_4$/C-G and $Li_3VO_4$/C -T composites, the stepped combustion of carbonaceous material can be attributed to the realization of different reactions (e.g., carbonization of glucose and tartaric acid, burning of amorphous carbon and the rest of carbon component, graphitization). In contrast, for $Li_3VO_4$/C-M composite the data imply only one intensive peak around 405 ºC which is associated with removal of $CO_2$ gases upon heating without visible satellite peaks attributed to multistage carbon oxidation. Based on the assumptions made above, the calculated carbon contents in $Li_3VO_4$/C-G, $Li_3VO_4$/C-M and $Li_3VO_4$/C-T composites are 25.6, 13.7, and 14.1 wt%, respectively (see Table 1). According to the CA, the bare $Li_3VO_4$ sample contains no carbon.

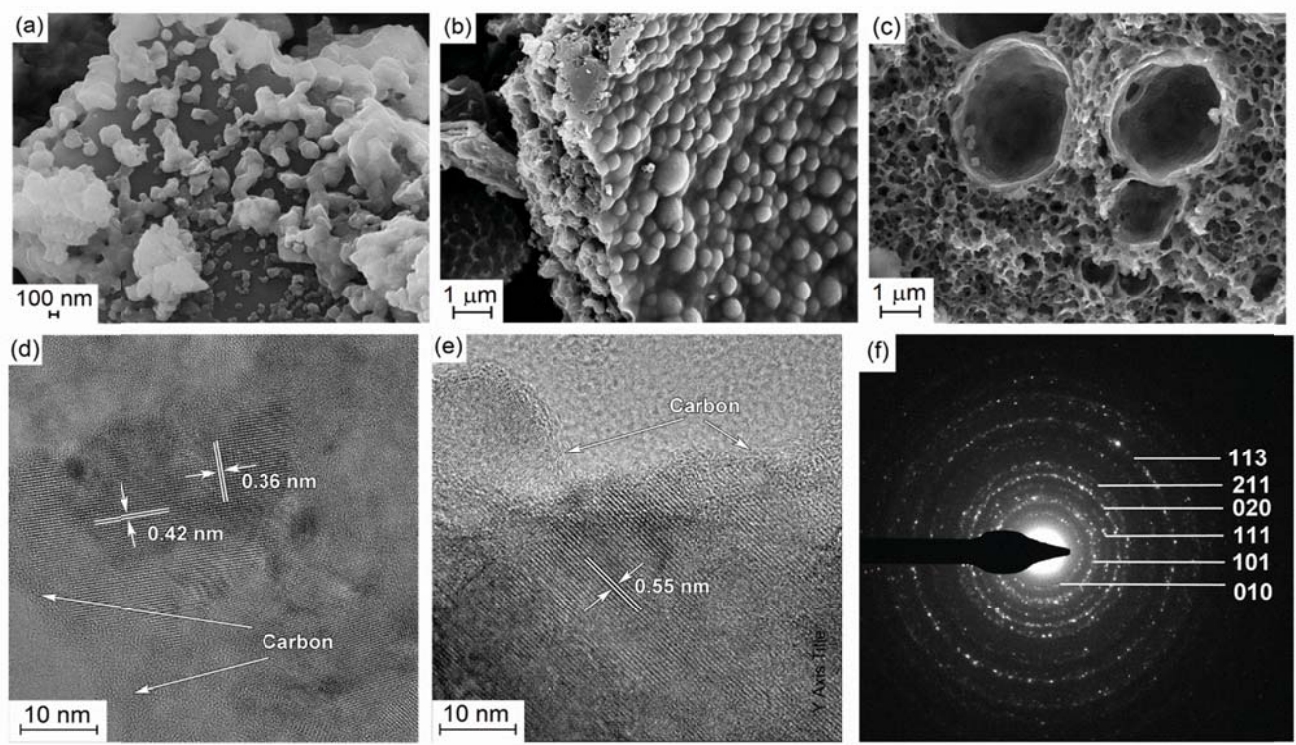

As can be seen from the SEM images (Fig. 3 and Fig. S1), the carbon source added during synthesis has a pronounced influence on the morphology of the resulting product. There are clear differences between $Li_3VO_4$/C-G, synthesized with glucose, and both samples which have been synthesized with carboxylic acids, i.e., $Li_3VO_4$/C-M and $Li_3VO_4$/C-T. $Li_3VO_4$/C-G composite consists of irregularly shaped particles of several tens of micrometers decorated with smaller particles which sizes are below 1 μm (Fig. 3a). In contrast, the powders $Li_3VO_4$/C-M and $Li_3VO_4$/C-T are composed of large sponge-like particles with partly smooth and partly mesoporous surface (Fig. 3b-c). In addition, micrometer sized round hollows are visible. The bare $Li_3VO_4$ sample consists of primary particles with sizes between 100 nanometers and few micrometers that are arranged to secondary particles of several tens of micrometers (Fig. S1).

A high resolution TEM (HRTEM) was used for further study the microstructure of the $Li_3VO_4$/C composites. Figures 3d-e show lattice fringes with interplanar spacing of 0.36, 0.42, and 0.55 nm, assigning to the (011), (110), and (010) planes of orthorhombic $Li_3VO_4$, respectively. Amorphous carbon layers with thickness of about 10 and 1 nm are observed on the surface of $Li_3VO_4$/C-G and $Li_3VO_4$/C-T. The results show that the thickness of the carbon layer covering the $Li_3VO_4$ particles is greater when glucose is used as carbon source as compared to tartaric acid. This observation complies with the results of thermogravimetric analysis. The selected-area electron diffraction (SAED) pattern of $Li_3VO_4$/C-T (Fig. 3f) indicate the polycrystalline nature of $Li_3VO_4$ with diffraction rings associated with the (010), (101), (111), (020), (211), and (113) crystal planes from inner to exterior, respectively.

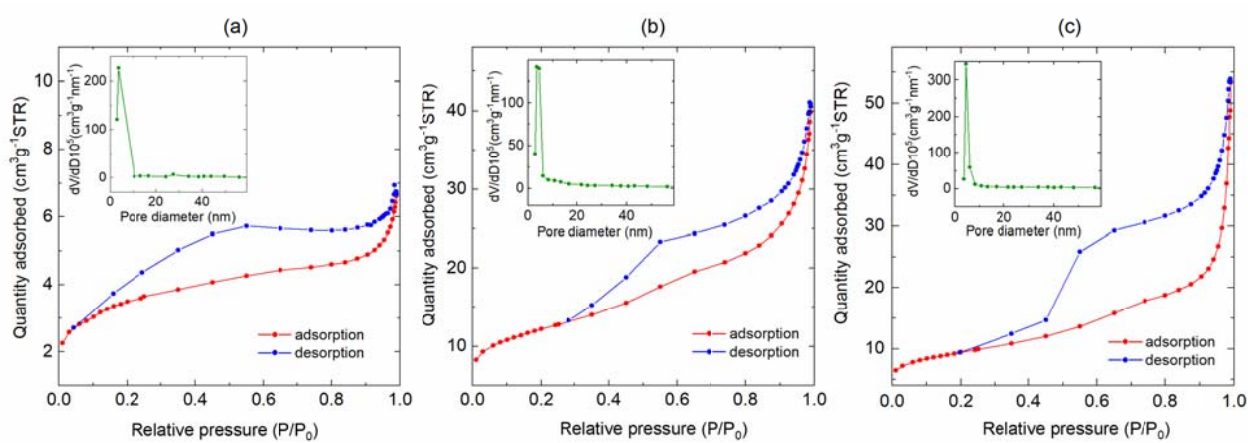

Fig. 4. Nitrogen sorption isotherms and corresponding pore-size distributions of (a) $Li_3VO_4$/C-G, (b) $Li_3VO_4$/C-M, and (c) $Li_3VO_4$/C-T composites.

Nitrogen sorption isotherms were generated to investigate the BET surface area and the porous structure of the $Li_3VO_4$/C composites (Fig. 4). All samples display a type-IV isotherm, based on the IUPAC classification, with an H3 hysteresis loop [41]. The BET specific surface area of the $Li_3VO_4$/C composites are displayed in Table 1 and are in the range of several 10 $m^2$ $g^{-1}$. In comparison with pure $Li_3VO_4$ [42], the BET surface area of the $Li_3VO_4$/C composites is rather high due to the presence of mesopores. Among the here reported composites, $Li_3VO_4$/C-M and $Li_3VO_4$/C-T show distinctly higher BET surfaces than the one produced by virtue of glucose. This is attributed to a highly porous structure and the presence of smaller particles as seen in the SEM images (Fig. 3b-c). The pore-size distribution curves demonstrate that the $Li_3VO_4$/C composites exhibit mesopores with a narrow pore size distribution of about 4 nm.

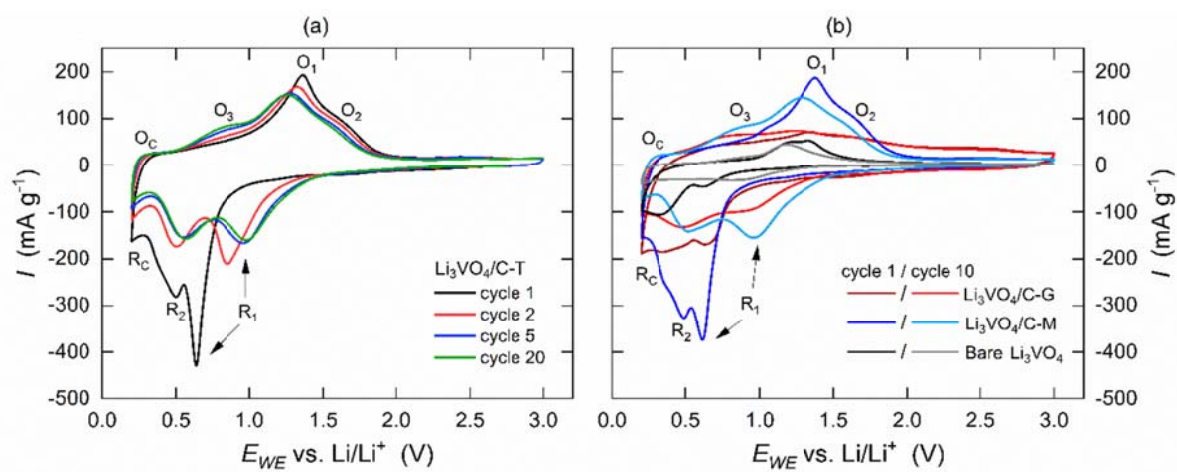

Fig. 5. CV curves recorded at a scan rate of 0.1 mV $s^{-1}$ of $Li_3VO_4$/C-T (a) and for comparison the first and tenth cycle of the different composites $Li_3VO_4$/C-G, $Li_3VO_4$/C-M, and bare $Li_3VO_4$.

In order to study the effects of various material parameters on the electrochemical properties of $Li_3VO_4$/C, cyclic voltammetry (CV) and galvanostatic cycling with potential limitation in the range 0.2 – 3 V are used. In addition, studies on bare $Li_3VO_4$ provide additional information about the effect of the carbon composite. In Fig. 5 the cyclic voltammograms of the $Li_3VO_4$/C composites recorded at a scan rate of 0.1 mV s$^{-1}$ are shown. Exemplarily, Fig 5a shows the first, second, fifth and twentieth cycle of the CV curves of $Li_3VO_4$/C-T. In the first reductive half-cycle two peaks at 0.64 V (R1) and 0.50 V (R2) occur that can be ascribed to the intercalation of $Li^+$ into $Li_3VO_4$ [43]. Additionally, in the first cycle the formation of the solid electrolyte interface (SEI) contributes in the same voltage range [6]. The lithiation and delithiation of the contained carbon gives rise to the redox peaks Rc and Oc at the lower cycling limit 0.2 V [44]. In the first oxidative scan two peaks at 1.36 V (O1) and 1.6 V (O2) are observable that correspond to the delithiation processes in $Li_3VO_4$ [43]. Upon further cycling the reduction peaks R2 and R3 are notably enlarged and shifted to higher potentials up to the 5th cycle. In oxidative sweep, besides a shift of the peaks O1 and O2 to lower potentials up to the 5th cycle, a third broad peak appears around 0.8 V (O3). Similar observations in the literature were associated with structural changes [33]. Iwama et al. [45] and Zhou et al. [43] showed that during the first lithiation there is an irreversible structure transformation. For comparison of the various $Li_3VO_4$/C composites reported at hand, the first and tenth cycle of the CV curves are shown in Fig. 5b. Despite differences in electrochemical activity between $Li_3VO_4$/C-G synthesized with glucose as carbon source as compared to $Li_3VO_4$/C-M and $Li_3VO_4$/C-T, synthesized with carboxylic acids, the courses of the cyclic voltammograms are almost similar. The electrochemical activity of $Li_3VO_4$/C-G is overall lower but extends over a wider potential range. For the reductive scan it ranges to lower potentials and for oxidative scans to higher potentials. However, specifically, all CV curves of the various composites exhibit the same general redox features described above for $Li_3VO_4$/C-T (Fig. 5a). The same applies to bare $Li_3VO_4$ which, however, features the lowest electrochemical activity.

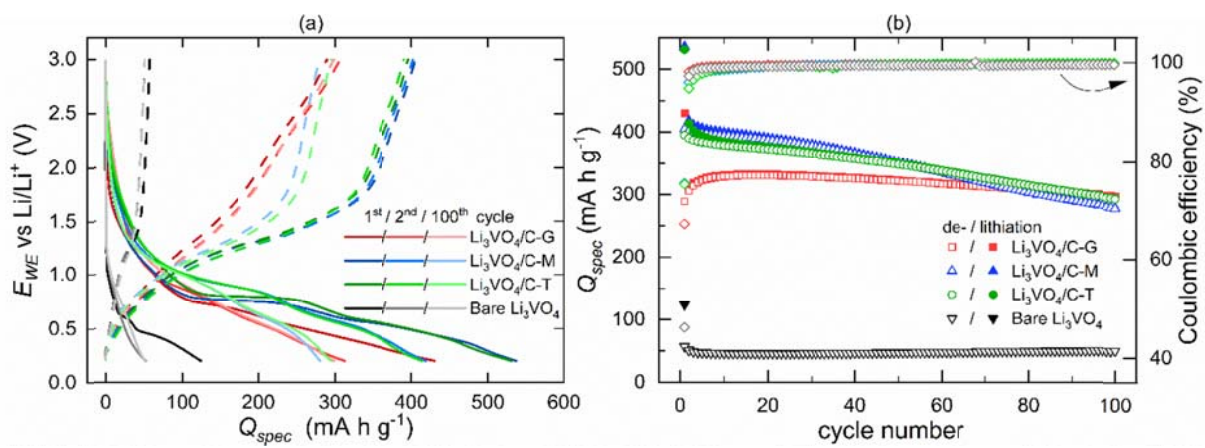

Fig. 6. Potential profiles of the first, second and hundredth cycle (a) and specific dis-/charge capacities and coulombic efficiencies (b) during galvanostatic cycling at 100 mA g$^{-1}$ of Li$_3$VO$_4$/C-G, Li$_3$VO$_4$/C-M, and Li$_3$VO$_4$/C-T composites as well as bare Li$_3$VO$_4$.

In Fig. 6 the long-term cycling performances of the three different composites during galvanostatic cycling at 100 mA g$^{-1}$ are shown. The potential profiles (Fig. 6a) are consistent with the results obtained from cyclic voltammetry (Fig. 5) and all redox features discussed above can be observed. In the first cycle the dis-/charge capacities are 430/289 mAh g$^{-1}$, 537/405 mAh g$^{-1}$, and 531/396 mAh g$^{-1}$ for Li$_3$VO$_4$/C-G, Li$_3$VO$_4$/C-M, and Li$_3$VO$_4$/C-T, respectively (Fig. 6b). The huge irreversible contribution in the first half cycle is caused by the SEI formation and structural changes [9,27,43]. Except for the first cycles, all samples display high coulombic efficiencies. As of the fifth cycle, the coulombic efficiency of Li$_3$VO$_4$/C-G is over 99%. By comparison, the composites Li$_3$VO$_4$/C-M and Li$_3$VO$_4$/C-T show lower values in the initial 20 cycles and exceed 99% in cycle 17 and cycle 13, respectively. There are clear differences in the cycle performance between the sample Li$_3$VO$_4$/C-G synthesized with glucose and the two composites synthesized with carboxylic acids, Li$_3$VO$_4$/C-M and Li$_3$VO$_4$/C-T. In the initial 50 cycles, the latter two reach significant higher capacities than Li$_3$VO$_4$/C-G. It is noticeable, that with measured values of about 400 mAh g$^{-1}$ the full theoretical capacity for an insertion of 2 Li$^+$/f.u. is gained. Up to the hundredth cycles the discharge capacities continuously decrease to 281 mAh g$^{-1}$ (Li$_3$VO$_4$/C-M) or rather 295 mAh g$^{-1}$ (Li$_3$VO$_4$/C-T) corresponding to 67 % and 71 % relative to the second cycle. In contrast, Li$_3$VO$_4$/C-G reaches indeed only a maximal discharge capacity of 333 mAh g$^{-1}$ in cycle 21 but it exhibits an excellent cycle stability. After 100 cycles a discharge capacity of 299 mAh g$^{-1}$ is measured, which corresponds to a retention of 96 % relative to the second cycle. The initial superior electrochemical activity of the samples Li$_3$VO$_4$/C-M and Li$_3$VO$_4$/C-T can be explained by their larger surface area compared to Li$_3$VO$_4$/C-G (Table 1). It provides an enhanced Li-ion diffusion and a high contact area with the carbon, which offers favorable electron transfer [28,40,42]. However, on the other side, the larger exposed electrode surfaces yield enhanced degradation due to corrosion and electrolyte dissolution, thus resulting in worse cycling stability [27,28]. Moreover, the cycle life is strongly influenced by the depth of discharge. The storage of a larger amount of Li-ions is accompanied with more serious structural damages. Crystal distortion and volume expansion by Li-ion insertion generate stress, which can lead to particle cracking. Such cracking in the electrode material has been considered to be one of the major mechanism for performance degradation [27,46]. This is in accordance with the observations made for the composite Li$_3$VO$_4$/C-G. It has the smallest surface area (Table 1), reaches initially the lowest capacity but is least affected by the degradation mechanisms discussed above. In addition, the high carbon content of Li$_3$VO$_4$/C-G is possibly another important factor that explains the enhanced cycling stability. The carbon plays an important role in inhibiting the degradation processes. On the one hand it protects the surface of the Li$_3$VO$_4$ particles for side reactions at the electrode-electrolyte interface [9,26–28]. Secondly, the carbon matrix can serve as a buffer for the volume changes during electrochemical cycling and prevents the electrode material from pulverization and aggregation [7,26]. A high and uniform distribution of the carbon in the electrode reduces the possibility of electrical contact loss and the formation of inactive regions in the material leading to performance degradation. The comparison of the Li$_3$VO$_4$/C composites with bare Li$_3$VO$_4$ clearly demonstrates the beneficial effect of the carbon composite. The specific capacity of bare Li$_3$VO$_4$ is far below those of the composite materials. In the first cycle the dis-/charge capacities are only 124/57 mAh g$^{-1}$ exhibiting low first columbic efficiency of 46 % due to side reactions at the unprotected Li$_3$VO$_4$/electrolyte interface [9]. The low specific capacity of bare Li$_3$VO$_4$ can be explained by limited access and inactive regions of the material due to its low electronic conductivity. *Ex-situ* XRD studies (Fig. S2) indeed suggest that a large part of Li$_3$VO$_4$ is not involved in the lithiation process. While after discharging to 0.2 V at 20 mA g$^{-1}$ the XRD pattern of Li$_3$VO$_4$/C-T electrode exhibits no more Bragg peaks associated with Li$_3$VO$_4$, for the discharged bare Li$_3$VO$_4$ electrode just a slight decrease in the intensity of the Li$_3$VO$_4$ Bragg peaks can be observed.

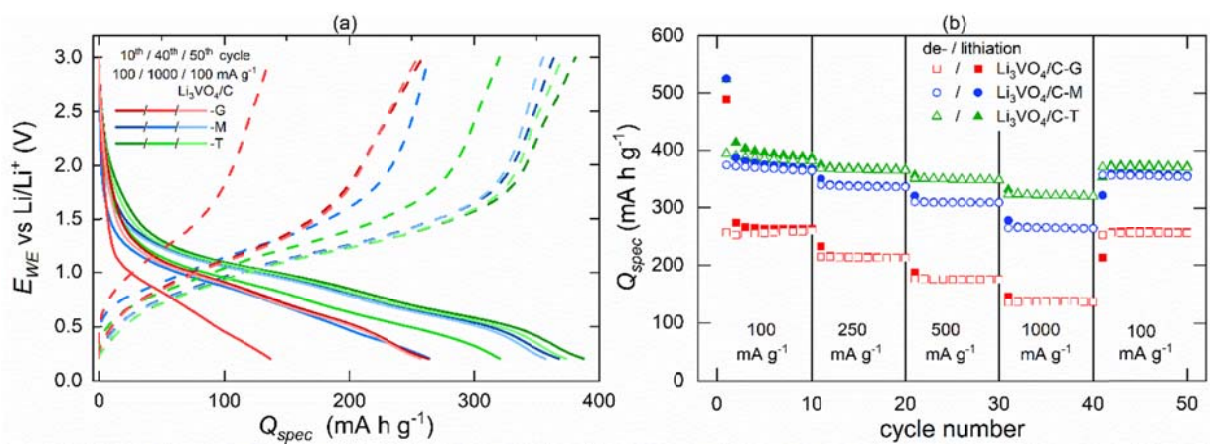

Fig. 7. Specific dis-/charge capacities (a) and potential profiles (b) for rate capability test with different cycling rates between 100 and 1000 mA g$^{-1}$ of Li$_3$VO$_4$/C-G, Li$_3$VO$_4$/C-M and Li$_3$VO$_4$/C-T composites.

To test the rate capability the samples were cycled with rates between 100 and 1000 mA g$^{-1}$. The specific dis-/charge capacities are shown in Fig. 7a and the potential profiles of cycle 10 at 100 mA g$^{-1}$, cycle 40 at 1000 mA g$^{-1}$ and cycle 50 again at 100 mA $g^{-1}$ in Fig. 7b. The composite $Li_3VO_4$/C-T reaches even at a high current density of 1000 mA $g^{-1}$ an excellent capacity of about 325 mAh $g^{-1}$. Furthermore, for all samples the charge capacity can regain to about 97 % when the current density returns to 100 mA $g^{-1}$. Obviously, high cycling rates do not cause irreversible damage detrimental for the electrochemical processes. In fact, the cycle stability is even worse for a constant low current density of 100 mA $g^{-1}$ (Fig. 6a). During the long-term cyclability test the capacity loss between cycle 10 and 41 of $Li_3VO_4$/C-M is 8 % and 6 % for $Li_3VO_4$/C-T. We attribute this to higher degradation effects of a huge discharge depth in case of low cycling rates. As demonstrated by the potential profiles, the worse rate capability of the sample $Li_3VO_4$/C-G and $Li_3VO_4$/C-M compared to $Li_3VO_4$/C-T can be explained by a larger resistance polarization. The potential gap between charge and discharge is significantly enlarged at a rate of 1000 mA $g^{-1}$ due to kinetic effects.

In table 2 the electrochemical performances of various $Li_3VO_4$-based anodes reported in literature are collected. In comparison to bare $Li_3VO_4$ the carbon composites exhibit enhanced $Li^+$ storage properties, especially in terms of electrochemical activity. The differences of the various composites can be traced back to differences in the carbon component on the one hand and morphological features on the other hand. In addition, the mere content of carbon and its structure plays an important role.

Table 2

Summary of the electrochemical performance of $Li_3VO_4$-based electrodes prepared via different synthesis methods.

| Synthesis method | Composite | Current 1C ≈ 394 mA $g^{-1}$ (mA $g^{-1}$) | 1$^{st}$ charge capacity (mAh $g^{-1}$) | Specific capacity (mAh $g^{-1}$) | Ref |
|---|---|---|---|---|---|

|  |  |  |  | /cycle |  |
|---|---|---|---|---|---|
| Microwave-assisted hydrothermal | $Li_3VO_4$/C nanoparticles | 100 | 163 | 104/100 | [42] |
| Hydrothermal* | $Li_3VO_4$/C nanoparticles | 150 | 594 | 542/300 | [33] |
| Solid-state combining with CVD* | $Li_3VO_4$/C nanospheres | 20 | 295** | 245/50 | [28] |
| Spray-drying* | $Li_3VO_4$/C hollow spheres | 80 | 429 | 400*/100 | [16] |
| Electrospinning* | $Li_3VO_4$/C nanofibers | 40 | 451 | 394/100 | [26] |
| Sol-gel* | $Li_3VO_4$/C nanoparticles | 100 | 491 | 408/100 | [19] |
| Citric acid-assisted sol-gel* | $Li_3VO_4$/C nanoparticles | 0.8C | 400** | 363/40 | [34] |
| Glucose-assisted sol-gel* | $Li_3VO_4$/C microparticles | 100 | 289 | 299/100 | This work |
| Malic acid-assisted sol-gel* | $Li_3VO_4$/C mesoporous particles | 100 | 405 | 281/100 | This work |

Synthesis methods marked with an asterisk include post-synthesis heat treatment.

Values marked with a double asterisk (**) are estimated from the graphs.

**Conclusion**

In summary, various $Li_3VO_4$/C composites with differences in morphology were prepared by a sol-gel method and subsequent annealing using different carbon sources. It is demonstrated that the electrochemical properties of $Li_3VO_4$/C as anode material for Li-ion batteries depend on morphology, surface texture and carbon content. In the initial 50 cycles, the mesoporous composites $Li_3VO_4$/C-M and $Li_3VO_4$/C-T synthesized with carboxylic acids excels in enhanced reversible electrochemical capacity which we attribute to their high surface area. However, both materials display slightly stronger fading effects which can be as well ascribed to adverse effects of the high surface area as a to huge charge depth upon the beginning cycles. While displaying slightly smaller initial reversible capacity, $Li_3VO_4$/C-G synthesized with glucose exhibits an excellent cycling stability. It is much less affected by the degradation mechanisms due to the smaller surface area and higher carbon content. Comparison with bare $Li_3VO_4$ clearly demonstrates the beneficial effect of carbon composites on the electrochemical performance. The data show that it significantly enhances specific capacity and improves first coulombic efficiency. A promising approach for further optimization may be the complete encapsulation of single $Li_3VO_4$ particles with carbon to protect the surface for side reactions at the electrode-electrolyte interface and suppress the pulverization upon cycling.

**Acknowledgements**

This work was supported by the Deutsche Forschungsgemeinschaft through project KL 1824/14-1. G.Z. acknowledges support of the state order via the Ministry of Science and High Education of Russia (Theme no. AAAA-A19-119031890025-9). E.T. and R.K acknowledges support by the BMWi through project 03ET6095C (HiKoMat). The authors thank I. Glass for experimental support.